\documentclass[11pt,preprint]{aastex}
\pdfoutput=1
\newcommand{\simgt}{\;\hbox{\rlap{\raise 0.425ex\hbox{$>$}}\lower 0.65ex\hbox{$\sim$}}\;}
\newcommand{\simlt}{\;\hbox{\rlap{\raise 0.425ex\hbox{$<$}}\lower 0.65ex\hbox{$\sim$}}\;}

\begin{document}

\title{Statistical mechanics of collisionless orbits. III.
Comparison with N-body simulations}

\author{ 
Liliya L. R. Williams\altaffilmark{1},
Jens Hjorth\altaffilmark{2}
\, and \, Rados{\l}aw Wojtak\altaffilmark{2,3}
}

\altaffiltext{1}{Department of Astronomy, University of Minnesota, 
116 Church Street SE, Minneapolis, MN 55455; llrw@astro.umn.edu}
\altaffiltext{2}{Dark Cosmology Centre, Niels Bohr Institute, 
University of Copenhagen, Juliane Maries Vej 30, DK-2100 Copenhagen \O, 
Denmark; jens@dark-cosmology.dk}
\altaffiltext{3}{Nicolaus Copernicus Astronomical Center, Bartycka 18, 
00-716 Warsaw, Poland; wojtak@camk.edu.pl} 

\begin{abstract}
We compare the DARKexp differential energy distribution, 
$N(\varepsilon) \propto \exp(\phi_0-\varepsilon)-1$, obtained from statistical 
mechanical considerations, to the results of N-body simulations of dark matter 
halos.  We first demonstrate that if DARKexp halos had anisotropic 
velocity distributions similar to those of N-body simulated halos, their 
density and energy distributions could not be distinguished from those of 
isotropic DARKexp halos. We next carry out the comparison in two ways, 
using (1) the actual energy distribution extracted from simulations, and 
(2) N-body fitting formula for the density distribution as well as $N(E)$ 
computed from the density using the isotropic Eddington formula. Both the
methods independently agree that DARKexp $N(E)$ with $\phi_0\approx 4-5$ is 
an excellent match to N-body $N(E)$. Our results suggest (but do not prove) 
that statistical mechanical principles of maximum entropy can be used to 
explain the equilibrated final product of N-body simulations.
\end{abstract}

\section{Introduction}\label{intro}

In this series of papers we revisit the use of statistical mechanics to
address the structure of self-gravitating N-body systems, such as dark matter
halos. In Hjorth \& Williams (2010; Paper I) we showed that if it is correct 
to think of finite self-gravitating collisionless isotropic systems as the 
most probable configuration in the energy state space, then their energy 
distribution is given by a truncated exponential differential energy 
distribution, DARKexp; $N(\varepsilon) \propto \exp(\phi_0-\varepsilon)-1$, 
where $\varepsilon$ and $\phi_0$ are dimensionless energy and potential depth,
respectively. The resulting structures have central density cusps.

In Williams \& Hjorth (2010; Paper II) we used the Extended Secondary Infall Model 
(ESIM) to confirm that restricted dynamical evolution can drive a system to the 
DARKexp state. ESIM largely fulfills the stipulations imposed by DARKexp; it is 
spherically symmetric, and its dynamics make for efficient energy redistribution 
among particles, but does not allow the redistribution of angular momentum, which
is not addressed in deriving DARKexp.
ESIM generates a wide range of potential depths, as quantified by DARKexp's
sole parameter, $\phi_0$.  ESIM halos are not exactly isotropic, many show 
radial anisotropy, and some have tangential anisotropies. The anisotropy
profiles vary between halos. These variations appear to be random; with no 
consistency, or `universality' between halos. 

In contrast, N-body halos do show well defined anisotropy, which appears to be 
as universal as the density profile \citep{hm06,h10}.  The halo
centers are isotropic but the outer parts have significant radial anisotropy. 
It is not surprising that N-body and ESIM halos have different anisotropy
characteristics. ESIM dynamics include radial forces and accelerations, but
no tangential accelerations (only velocities). Simulations, on the other hand, 
operate in full three dimensions, so tangential forces can redistribute 
particles' angular momenta and establish a universal anisotropy profile. 
Therefore it is possible that the presence of the well defined, universal 
anisotropy profile of N-body halos indicates an important element of the 
equilibrium structure of simulated halos, not accounted for in DARKexp models.

A full theory for the distribution of particles in energy-angular momentum space 
using the maximum entropy principle is yet to be developed. In the meantime, 
we ask if we can limit ourselves to comparing energy distributions only, without 
regard to non-zero anisotropy. Since whatever produces the anisotropy profile in
simulated halos may also effect the energy distribution, it is not immediately 
obvious that DARKexp $N(E)$ can be compared to that of N-body halos. To proceed 
one must first determine what influence the anisotropy has on DARKexp $N(E)$ and 
density profile.  If the anisotropy of N-body halos is such that it has no 
effect on $N(E)$ and $\rho(r)$, then one can compare the DARKexp and N-body 
$N(E)$ without worrying that anisotropy---and by implication whatever physics 
is present in simulations but was neglected in deriving DARKexp---will 
invalidate the comparison.

In Section~\ref{fullN} we show that if anisotropy is similar to what the N-body 
simulations produce, then $N(E)$ and $\rho(r)$ are unchanged compared to their
isotropic counterparts. Therefore we argue that a comparison between DARKexp and 
N-body $N(E)$ can be justifiably made, and carry it out in two ways, as
described in Sections~\ref{compN} and \ref{compA}.

\section{$N(E,L)$ distributions of DARKexp models}\label{fullN}

\subsection{Generating self-consistent sets of $N(E,L)$, $\rho(r)$ and $\beta(r)$}\label{meth}

A geometrically spherically symmetric non-rotating system is fully described 
by the distribution of its particles in the two dimensional plane of energy, 
$E$ and angular momentum, $L$;~ $N(E,L)$. The range of energies is bounded on 
one end by the value of the deepest central potential, and the escape energy, 
$E=0$ on the other end. For every $E$ there is a maximum $L$ which corresponds 
to the circular orbit at that energy;  $L_{max}(E)$ is an upper envelope in the 
($E$, $L$) plane. The lower boundary is $L=0$, implying a radial orbit. 
An example of an $L$ vs.\ $E$ plot and the $L_{max}(E)$ envelope can be found 
in \cite{w08}, for the particular case of numerically generated universal halos. 

Even though the $N(E,L)$ distribution contains all the information about 
a system, there is no simple relation (that we know of) between it and
the corresponding density and anisotropy profiles. 

For the purposes of this work we need to check if a set of $N(E,L)$, $\rho(r)$, 
and anisotropy distributions is self-consistent (Sections~\ref{iso} and \ref{ani1}), 
or, generate self-consistent sets of these distributions (Section~\ref{ani2}).
Anisotropy is defined as usual, $\beta(r)=1-\frac{\sigma^2_\theta(r)}{\sigma^2_r(r)}$; 
in spherical systems the two tangential velocity dispersions are equal, 
$\sigma_\theta=\sigma_\phi$. We use the following procedure. 
Given input distributions $N(E,L)$ and $\rho(r)$ we numerically generate a halo 
by drawing particles from the $N(E,L)$ distribution and adding up, or superimposing 
their orbits. The orbit superposition is done as follows. 
The system's potential $\Phi(r)$ is calculated from the input density profile.
For each particle picked randomly from $N(E,L)$ we find its radial velocity, 
$v_{rad}(r)$, and apo- and  peri-centers using the energy equation, 
$E=\Phi(r)+\frac{1}{2}\Bigl[v_{tan}^2(r)+v_{rad}^2(r)\Bigr]$, where the 
angular momentum is related to the radially dependent tangential velocity 
through $v_{tan}(r)=L/r$. The density that a particular particle contributes at 
a given radius is proportional to the amount of time it spends there, i.e., 
$\propto {{v_{rad}^{-1}\;dr}/{\int_{r_{peri}}^{r_{apo}}v_{rad}^{-1}\;dr}}$.
Thus a halo is built up from its individual particles. If the density profile 
generated in this fashion matches the input $\rho(r)$, then the set of $\rho(r)$ 
and $\beta(r)$ (computed afterward from the radial and tangential velocity 
dispersions) represent a self-consistent solution for the input $N(E,L)$. 

In this work, the only input energy distributions and density profiles we
use are those of the isotropic DARKexp models. In Sections~\ref{iso} and ~\ref{ani1} 
we use the above procedure to determine if specific $L$-distributions are 
consistent with DARKexp. In Section~\ref{ani2} we seek, through trial-and-error, 
an $L$-distribution that produces a specific anisotropy profile. So,
if for a given $L$-distribution the input $\rho(r)$ and the output $\rho(r)$ from 
orbit superposition do not match then the $L$-distribution is rejected. (Note 
that acceptance/rejection is not done on individual particle basis; only whole 
halos are accepted or rejected.) The procedure is repeated until an acceptable 
$L$-distribution is found.

This method has some features in common with the Schwarzschild orbit superposition
method \citep{sch79}, where a self-consistent equilibrium halo is built up from its 
constituent orbits. There is no dynamical evolution.

\subsection{Anisotropy and the distribution of $L$}\label{betaL}

To generate halos, one needs an input $N(E,L)$. While DARKexp gives us the 
distribution in $E$, its says nothing about how orbits should be distributed in 
$L$ to produce a given $\beta(r)$. To guide us, we start with some well known 
results connecting the energy distribution and the distribution function, $f$. 
In general, 
\begin{equation}
N(E,L)\;dE\;dL=8\pi^2\; L\; f(E,L)\; T_r(E,L)\; dE\; dL,
\label{NE1}
\end{equation}
where $T_r$ is the radial period of orbits (for example, see Appendix A of \cite{w08}).
For isotropic systems, $f=f(E)$. 
\cite{h59} proved that only three types of potentials have $T_r$ that depends on 
$E$ alone: the point mass, the uniform density sphere, and the isochrone potential 
(see also \cite{e07}). It is then more instructive to write eq.~\ref{NE1} as
\begin{equation}
N(E,L^2)\;dE\; dL^2=4\pi^2\;f(E)\;T_r(E)\;dE\; dL^2.
\label{NE2}
\end{equation}
The right hand side---and hence the left hand side---has no explicit $L$-dependence, 
therefore the number of orbits at a given $E$, between two $L$ values is 
proportional to $dL^2$ , which means that for these three potentials 
isotropy implies a uniform distribution in $L^2$. 

With this result in mind, one can rewrite eq.~\ref{NE1} to describe systems where 
orbits are distributed uniformly in $L^\delta$,
\begin{equation}
N(E,L^\delta)\; dE\; dL^\delta = \frac{8\pi^2}{\delta}\;L^{2-\delta}\; f\; T_r\;dE\; dL^\delta.
\label{NE3}
\end{equation}
If the distribution in a given parameter is uniform, that means the function has no
explicit dependence on it. In the case of eq.~\ref{NE3}, the expression for 
$N(E,L^\delta)$ has no explicit $L^\delta$, or $L$ dependence. Further
assuming~ $T_r=T_r(E)$, implies that $L^{2-\delta}\;f\equiv h(E)$ is a function of $E$ only, 
and hence $f$ is separable; $f=h(E)L^{\delta-2}$. These are systems of constant anisotropy, 
$\beta=1-\delta/2$ \citep{h73}. A $\beta=0$ system is recovered for $\delta=2$. Even though 
$T_r$ is not strictly a function of $E$ only for a general system, it appears to be 
a very good approximation for many potentials. Hence a uniform distribution in $L^\delta$ 
can be used to approximate constant $\beta$ systems of arbitrary density profiles.

\subsection{Isotropic halos}\label{iso}

\begin{figure}[t]
\epsscale{0.95}
\plotone{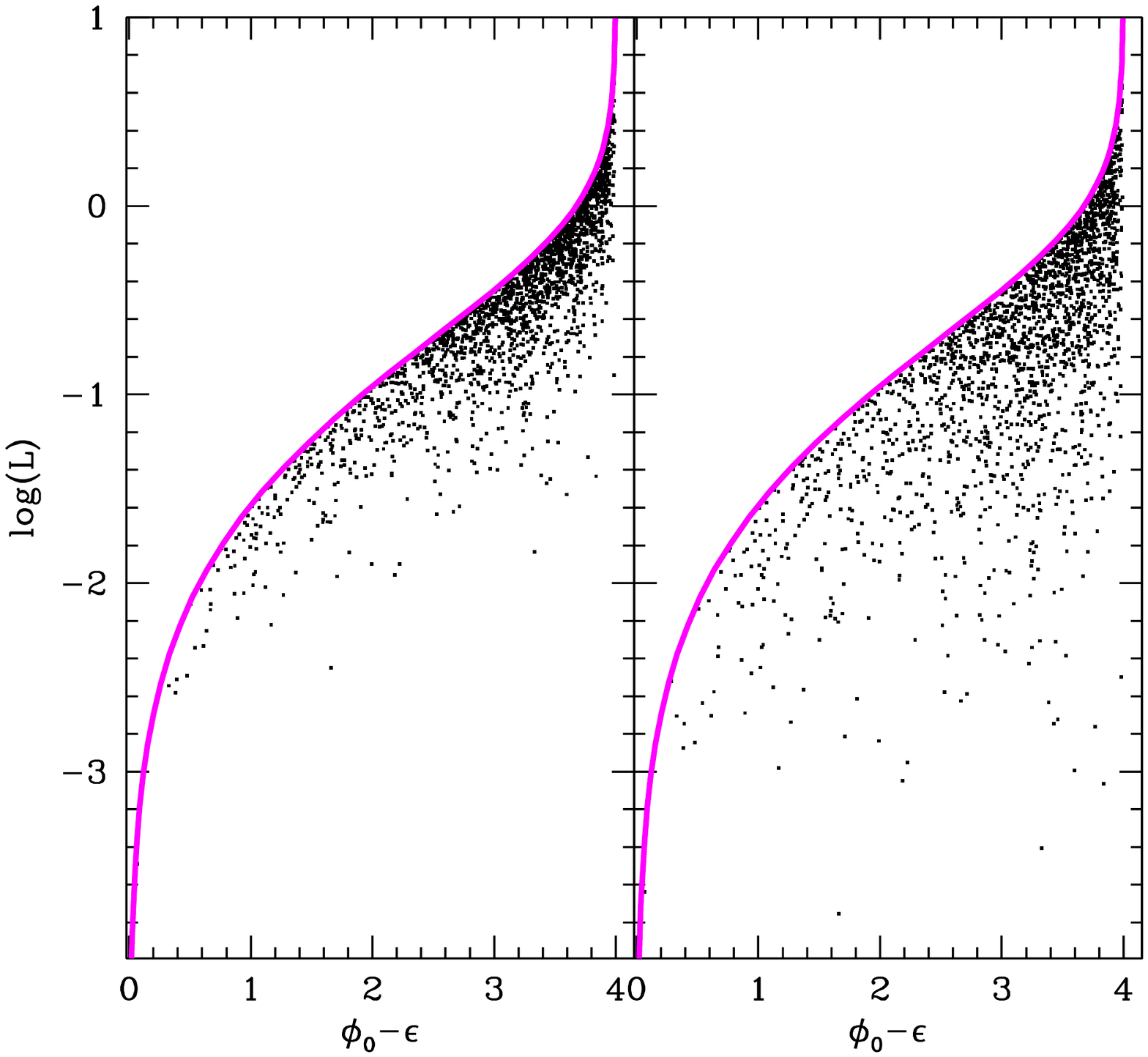}
\vskip-1.5in
\caption{The log of angular momentum vs.\ the energy of particles making up 
halos. The energy distribution is the DARKexp $N(\varepsilon)$, with $\phi_0=4$ 
and particles shown are a realization of this distribution. The magenta line
is the envelope corresponding to circular orbits, $L_{max}(E)$.
{\it Left panel:} 
For a given $E$, the distribution in $L^2$ is uniform between 0 and $L_{max}^2(E)$.
This gives an approximately isotropic system (Section~\ref{betaL}; Figure \ref{denbetaD}).
{\it Right panel:} 
For a given $E$, the distribution in $L$ is uniform between 0 and $L_{max}(E)$.
This gives constant $\beta(r)\approx 0.5$ (Section~\ref{betaL}; Figure \ref{denbetaB}).}
\label{EJ}
\end{figure}

\begin{figure}[t]
\epsscale{0.95}
\plotone{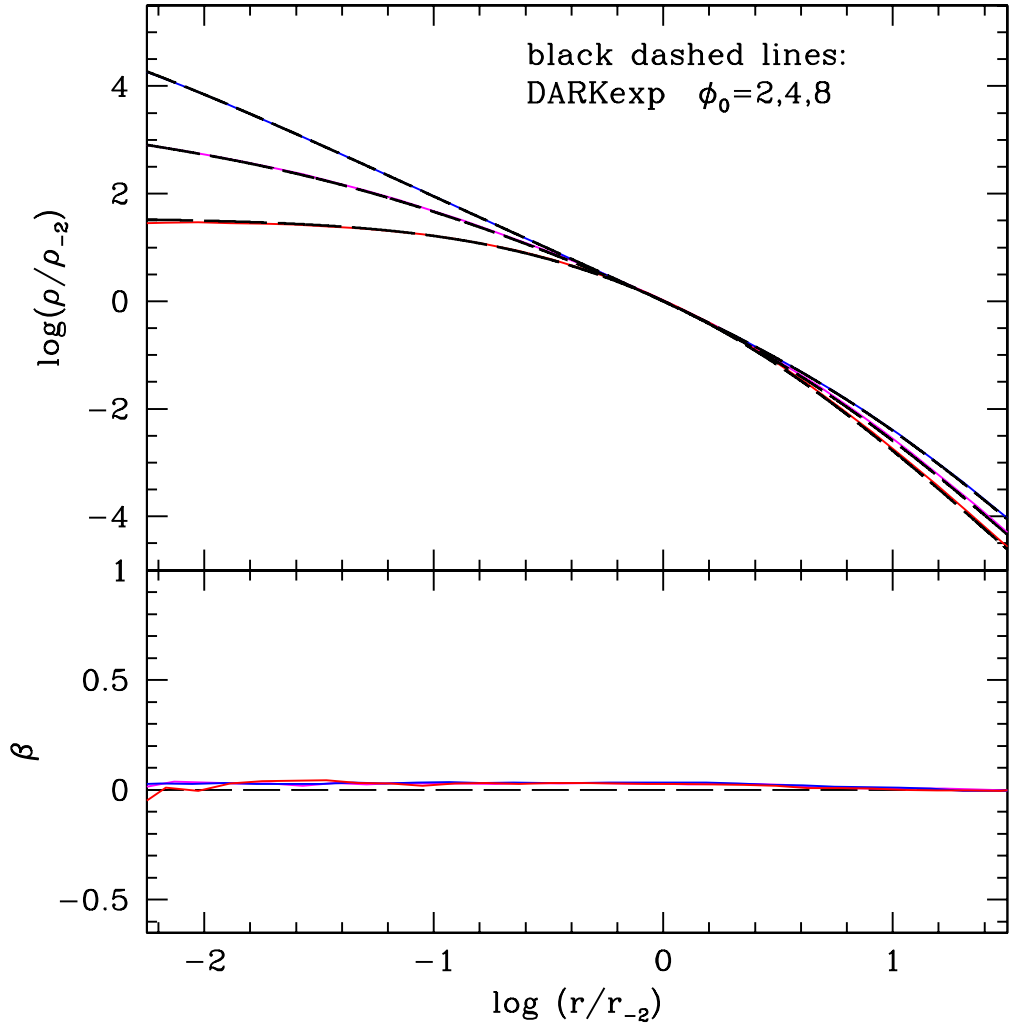}
\vskip-1.5in
\caption{Density and anisotropy profiles for DARKexp models (dashed black curves), 
and halos generated using the method described in Section~\ref{meth} (solid color curves).
To achieve isotropy for the red, magenta, and blue solid curve halos we used 
a uniform distribution in $L^2$, illustrated in the left panel of Figure~\ref{EJ}.} 
\label{denbetaD}
\end{figure}

In this section we generate $N(E,L)$ distributions for isotropic DARKexp models.
Although we know from \cite{h59} that a strictly uniform distribution in $L^\delta$
with $\delta=2$  cannot produce isotropic systems for DARKexp, we nevertheless try a
uniform distribution of particles in $L^2$. This is shown as black points in the 
left panel of Figure~\ref{EJ}, for DARKexp $\phi_0=4$. The magenta curve represents the
upper boundary, $L_{max}(E)$.
We apply the orbit superposition procedure described above, with an isotropic 
DARKexp $N(E)$ and $\rho(r)$ as the input, shown as dashed black curves in the top 
panel of Figure~\ref{denbetaD}. The resulting density profiles are shown as the 
red, magenta and blue solid lines for $\phi_0=2,4,8$ DARKexp models, respectively.
These agree very well with the input $\rho(r)$, making this a self-consistent set 
of isotropic DARKexp halos. 

We conclude that isotropic DARKexp halos (with DARKexp form for the $N(E)$) are very 
well described by a uniform orbit distribution in $L^\delta$, where $\delta=2$.
The deviations from this, which we know have to be present,
appear to be smaller than the precision required in the present paper.

\subsection{Anisotropic halos; $\beta\approx 0.5$}\label{ani1}

\begin{figure}[t]
\epsscale{0.95}
\plotone{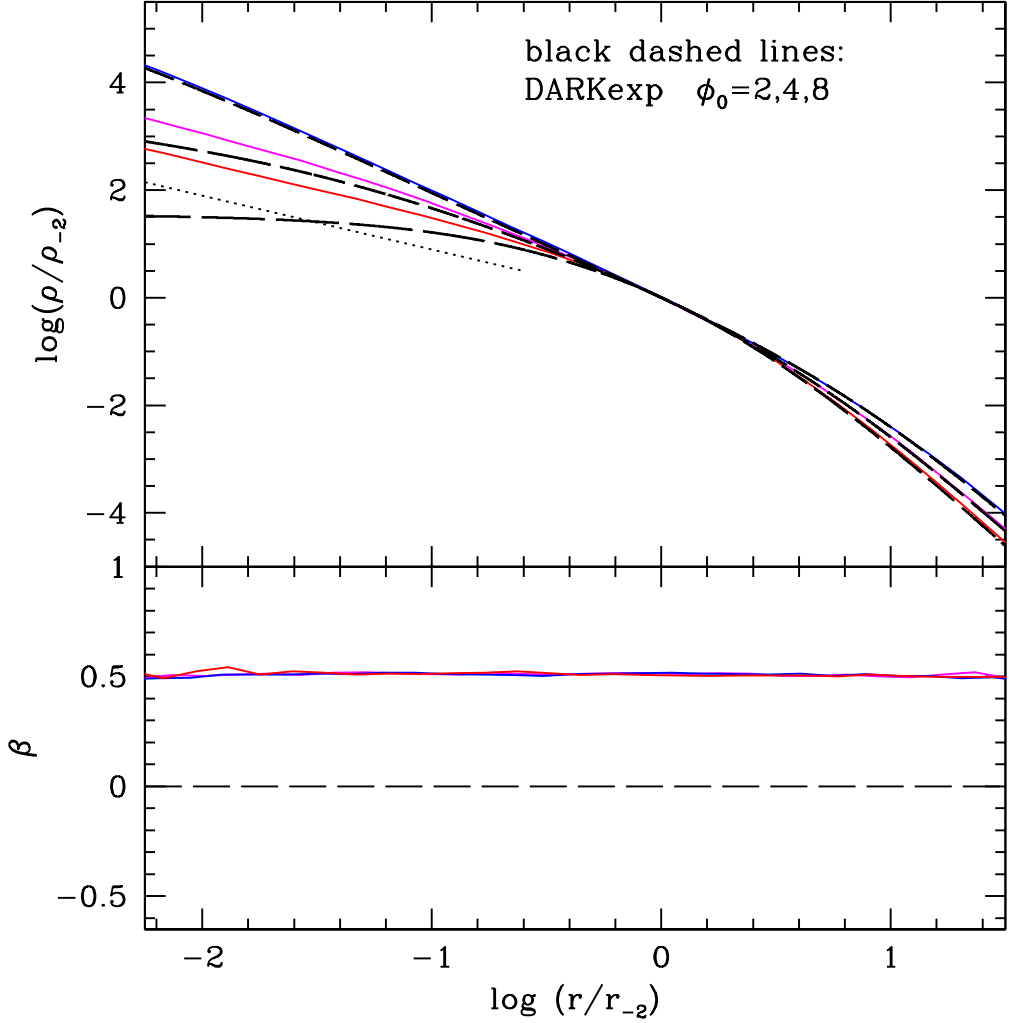}
\vskip-1.5in
\caption{Density and anisotropy profiles for DARKexp models (dashed black curves), 
and halos generated using the method described in Section~\ref{meth} (solid color curves).
At every $E$, the distribution in $L$ is uniform between 0 and $L_{max}(E)$; this 
distribution is illustrated in the right panel of Figure~\ref{EJ}, and produces halos
with $\beta\approx 0.5$. These systems obey the density-anisotropy relation, $\gamma\ge 2\beta$
since the density slope is never shallower than $\gamma=1$, shown as the thin dotted line
segment in the upper panel.} 
\label{denbetaB}
\end{figure}

For comparison, in the right panel of Figure~\ref{EJ} we show the case where, for 
any given $E$, the distribution is uniform in $L$, i.e. $\delta=1$, so by the arguments 
of Section~\ref{betaL} these systems should have (nearly) constant $\beta=0.5$.
The anisotropy profiles of the orbit superposition halos are presented in the 
bottom panel of Figure~\ref{denbetaB} as the red, magenta and blue solid curves 
for $\phi_0=2,4,8$ DARKexp models, respectively. They do, in fact, have $\beta\approx 0.5$. 

The DARKexp input density profiles are the as black dashed lines in the top panel, 
while the orbit superposition profiles are represented by red, magenta and blue lines. 
These do not match the input $\rho(r)$, and therefore DARKexp $N(E)$, which we used as
the input energy distribution, is not consistent 
with constant anisotropy as large as $\beta=0.5$.

A comment is in order. Inspection of Figure~\ref{denbetaB} shows that the systems obtained 
by orbit superposition obey the 
anisotropy-density inequality introduced by \cite{ae06} for the central halo regions,
and extended to apply to all radii by \cite{cm10}. It states that $\gamma\ge 2\beta$ 
(recall that $\rho (r) \propto r^{-\gamma(r)}$). Applied to our systems it means that if 
$\beta=0.5$, the density slope cannot be shallower than $\gamma=1$ anywhere in the system. 
This slope is represented by a dotted line segment in the upper panel of Figure~\ref{denbetaB}.
Our three density profiles are steeper than this at all radii. As a consequence of the 
anisotropy-density inequality, the oscillations in the density slope (Paper II) are erased, 
and the asymptotic slope is attained at much larger radii than in the isotropic models.

\subsection{Anisotropic halos; $\beta(r)\approx\beta_{\rm N-body\; sim}(r)$}\label{ani2}

Here we repeat the orbit superposition procedure but aim to generate anisotropy profiles 
similar to those of N-body simulated halos. Since these are not constant, 
a uniform distribution in $L^\delta$ will not work. Through trial and error we came up with a 
Gaussian distribution in $L^2$. The full distribution in the $E$ vs.\ $L^2$ plane is,
\begin{equation}
N(\epsilon,L^2)\;d\epsilon\;dL^2\propto N(\epsilon)\times
\exp\left\{-{\Bigl([L^2/L_{max}^2(\epsilon)]-[\frac{a_1}{a_2(\epsilon/\phi_0)}+a_2]\Bigr)^2}\Big/{(2a_3)}\right\}
\;d\epsilon\;dL^2,
\end{equation}
where $N(\epsilon)\propto\exp(\phi_0-\epsilon)-1$ is the DARKexp form. As in Papers 
I and II we use dimensionless energy and potential, $\epsilon=\beta_T E$, and 
$\phi_0=\beta_T \Phi_0$, where $\Phi_0$ is the system's potential depth, and $\beta_T$ 
is its (negative) inverse temperature.  The constants $(a_1,a_2,a_3)$ are
$(0.25,1.00,0.20)$ for $\phi_0=2$, 
$(0.25,0.75,0.15)$ for $\phi_0=4$, and
$(0.25,0.50,0.10)$ for $\phi_0=8$.

\begin{figure}[t]
\epsscale{0.95}
\plotone{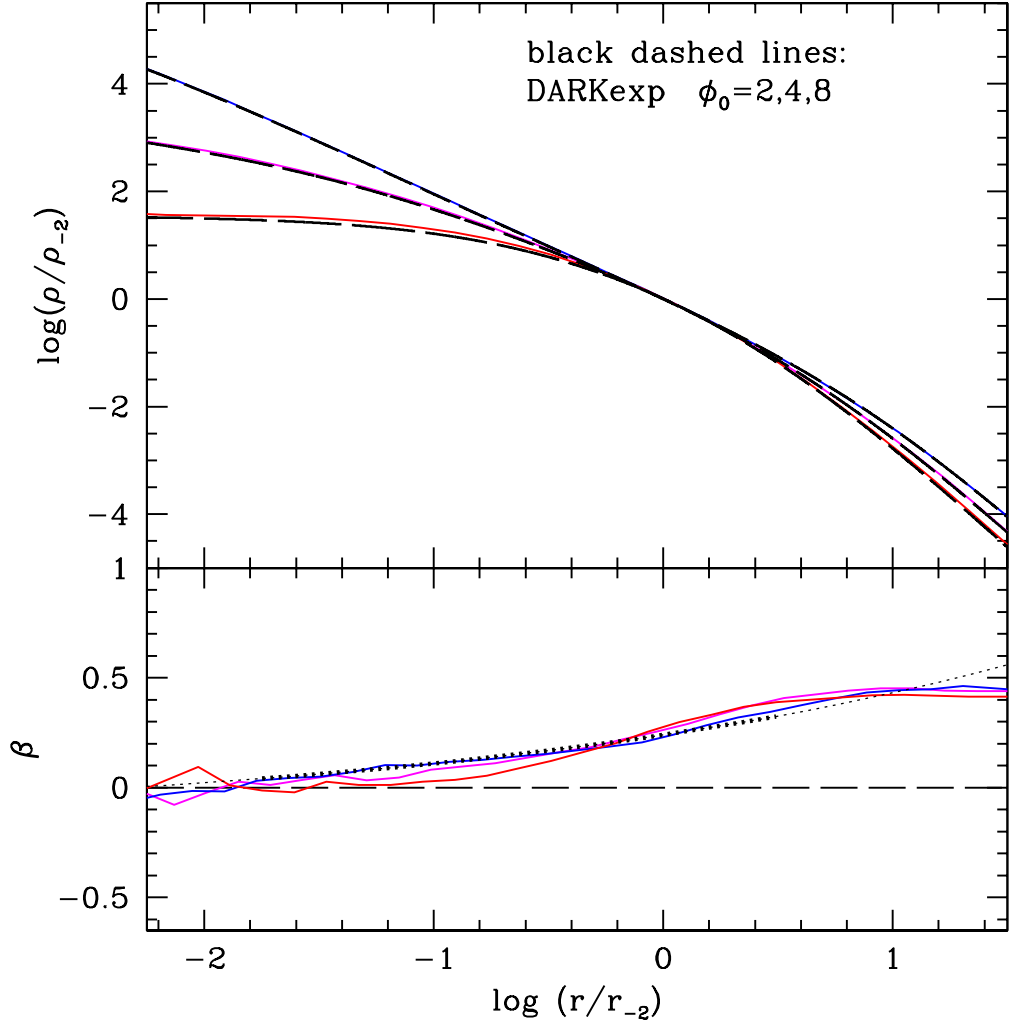}
\vskip-1.5in
\caption{Density and anisotropy profiles for DARKexp models (dashed black curves), 
and halos generated using the orbit superposition method described in Section~\ref{meth}. 
These density profiles (red, magenta, and blue curves) agree very well with the 
isotropic DARKexp $\rho(r)$. The distribution in the $L$ vs.\ $E$ plane (see Section~\ref{ani2})
was chosen such that the anisotropy profile would look similar to that of N-body 
halos. The latter is plotted (dotted line) using the Einasto profile 
$\gamma=2(r/r_{-2})^\alpha$ with $\alpha=0.17$ \citep{nav04} combined with the 
anisotropy-slope relation, $\gamma=-0.8-5\beta$ \citep{h10}. The thick portion of the 
dotted line represents the region where the $\gamma-\beta$ relation can be trusted.}
\label{denbetaC}
\end{figure}

The density and anisotropy profiles are shown as red, magenta and blue solid lines 
in the two panels of Figure~\ref{denbetaC}. The anisotropy profiles generated through 
orbit superposition (red, magenta and blue curves in the bottom panel) are quite 
close in shape to the universal anisotropy profile. The latter is plotted (dotted line)
using the Einasto profile $\gamma=2(r/r_{-2})^\alpha$ with $\alpha=0.17$ \citep{nav04} 
combined with the anisotropy-density slope relation, $\gamma=-0.8-5\beta$ \citep{h10}. 
The thick portion of the dotted line represents the region where the $\gamma-\beta$ 
relation can be trusted \citep{nav10}.

Each of the three orbit superposition density profiles (red, magenta, blue) matches 
its corresponding isotropic DARKexp profile (dashed black) very closely, which means 
that DARKexp $N(E)$ remains a good description for systems with anisotropies as large 
as those seen in simulations of universal halos. In other words, DARKexp $N(E)$ and 
density profiles originally derived for isotropic systems, are also consistent with 
systems whose $\beta(r)$ profiles are isotropic at the center and radially anisotropic 
closer to virial radii. We conclude that we can compare DARKexp $N(E)$ to that of 
N-body simulated halos, and safely ignore the anisotropy of the latter.

\section{Comparing DARKexp $N(E)$ to that of simulated halos}\label{comparison}

Having demonstrated that universal anisotropy is too small to have an effect on $N(E)$ 
we now compare the DARKexp energy distribution to that of simulated universal
halos, using two methods, described below.

\subsection{Using $N(E)$ of simulated halos}\label{compN}

Here we determine the differential energy distribution of dark matter particles 
residing inside the virial sphere of simulated halos. For this purpose we 
use a sample of 36 cluster-size relaxed halos extracted from a $z=0$ snapshot of a
$N$-body simulation of a standard $\Lambda$CDM cosmological model \citep[for details of 
the simulation and the halo catalogue see][]{w08}. This halo sample has already 
been used for calculating the distribution function and testing its phenomenological 
model with radially changing anisotropy \citep{w08}. Each halo contains from 
$5\times 10^{5}$ to $5\times 10^{6}$ particles inside its virial sphere defined 
in terms of the mean overdensity, $\langle \rho\rangle/\rho_{\rm c}\approx 100$, 
where $\rho_{\rm c}$ is the present critical density.

\begin{figure}[t]
\vskip-2.0in
\epsscale{0.95}
\plotone{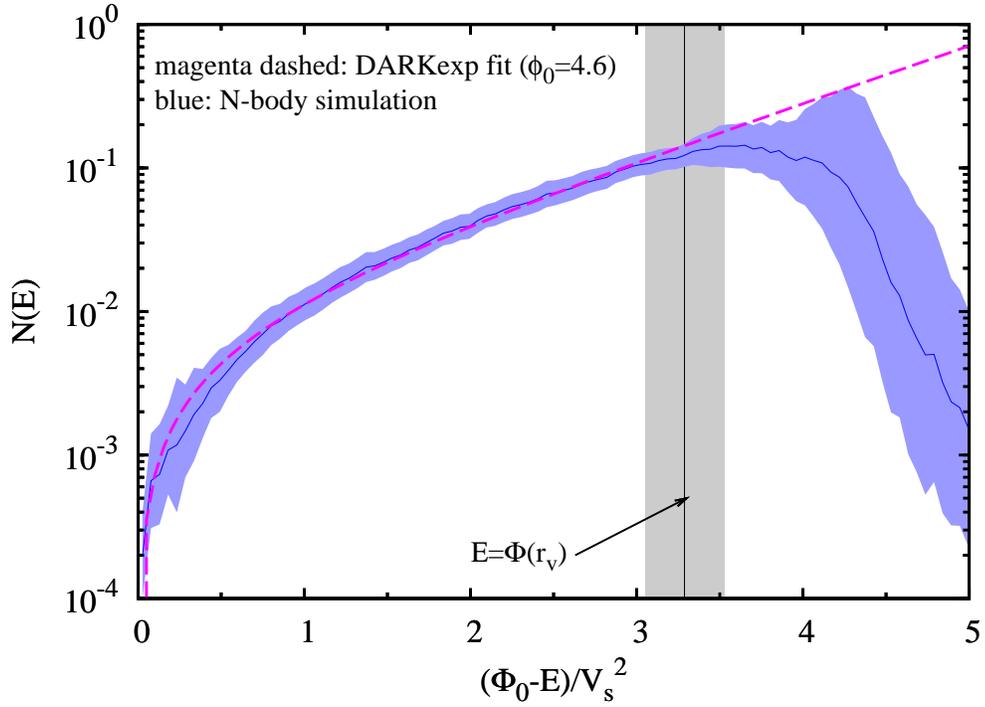}
\caption{Comparison between $N(E)$ from $N$-body simulation (blue curve) and the 
DARKexp fit with $\phi_{0}=4.6$ (dashed magenta curve). The blue curve and light blue 
area represent the median profile and the 1$\sigma$ scatter within the halo sample. The 
light gray area indicates the gravitational potential at the virial sphere 
(with the width corresponding to the 1$\sigma$ scatter within the halo sample and the black 
line showing the median value). It defines an upper limit of the relative energy for a set 
of particle orbits which are fully confined inside the virial sphere.}
\label{DE36halo}
\end{figure}

We calculate $N(E)$ for each halo independently by counting particles in energy 
bins. Then we combine all profiles into one and evaluate the median profile and the 
dispersion within the halo sample (blue curve and light blue area in Figure~\ref{DE36halo}). 
To combine the halos, we scale  the binding energies by $V_{\rm s}^{2}=GM(<r_{-2})/r_{-2}$ 
and the particle numbers by $N(<r_{-2})$, where $r_{-2}$ is the radius 
where $\gamma(r_{-2})=2$. These scalings preserve scale-free similarities of the 
phase-space properties of halos including the differential energy distribution itself. 
They also diminish the differences between $N(E)$ profiles of individual halos with 
varying virial masses and scale radii. The scale radii, $r_{-2}$, were obtained by 
fitting the NFW profile to the density profiles of simulated halos.  We emphasize that 
the results are independent of the assumption of a fitting formula and remain the same 
if the scale radii are calculated by finding a maximum of $\rho(r)r^{2}$, which is a 
fitting-independent method of measuring $r_{-2}$.

Although our analysis is restricted to the virialized parts of halos, we expect 
to encounter particle populations which are not fully equilibrated, e.g., particles 
which have only recently entered the virial sphere. In order to separate the energy 
range of these particles we indicate in Figure~\ref{DE36halo} the value of the 
gravitational potential at the virial radius $r_{\rm v}$ of all halos (see the black 
line and light gray rectangle showing the median value and the 1$\sigma$ scatter 
within the halo sample). All relative energies, $\Phi_{0}-E$, below this value are 
covered by particles moving along the orbits confined inside the virial sphere so that 
this energy range likely represents a fully equilibrated particle population. On the 
other hand, the upper energy range is partially populated by orbits extending beyond 
the virial sphere so that it likely includes a particle population which has not 
settled to equilibrium. It is worth noting that at the energy separating these two 
energy regimes, i.e., $E\approx \Phi(r_{\rm v})$, $N(E)$ starts deviating from an 
exponential growth predicted by the DARKexp model. At less bound energies, 
$E>\Phi(r_{\rm v})$, the presence of unequilibrated particle population manifests 
itself in an exponential cut-off of $N(E)$.

We fitted the median profile of $N(E)$ for $N$-body halos with the DARKexp model 
given by $N/N_{0}=\exp[-\beta_{T}(E-\Phi_{0})]-1$, parametrized by the normalization 
$N_{0}$ and the inverse temperature $\beta_{T}$. The fitting was restricted to the energy 
range associated with the fully equilibrated particle population, i.e., 
$\Phi_{0}-E\lesssim 3V_{s}^{2}$ (see Figure~\ref{DE36halo}).
It yields $\beta_{T}=0.91\,V_{\rm s}^{2}$, which leads to the dimensionless potential 
depth of the halos $\phi_{0}=4.6$. This best fit $\phi_{0}$ is consistent 
with the most favourable range of values resulting from comparing DARKexp model with 
the best analytical approximations of the universal density profile ($\phi_{0}\approx 4-5$; 
Section~\ref{compA}). Given this agreement in $\phi_0$, and the results in 
Figure~\ref{DE36halo} we conclude that the DARKexp fit (dashed magenta curve) 
is a very good match to the $N(E)$ profile from simulations.

\subsection{Using $\rho(r)$ fitted to simulated halos}\label{compA}

Several fitting formulae have been suggested to describe the density profiles
of N-body halos. The original double power-law NFW form was introduced by
\cite{nfw97} as the universal shape for cold dark matter halos. It was later 
pointed out that simulated halos are even better fit with Einasto profiles 
\citep{nav04,m06}. The S\&M profile \citep{sm09} was recently proposed as a
very good fit to the inner most regions, $r\simlt r_{-2}$.

\begin{figure}[t]
\epsscale{0.95}
\plotone{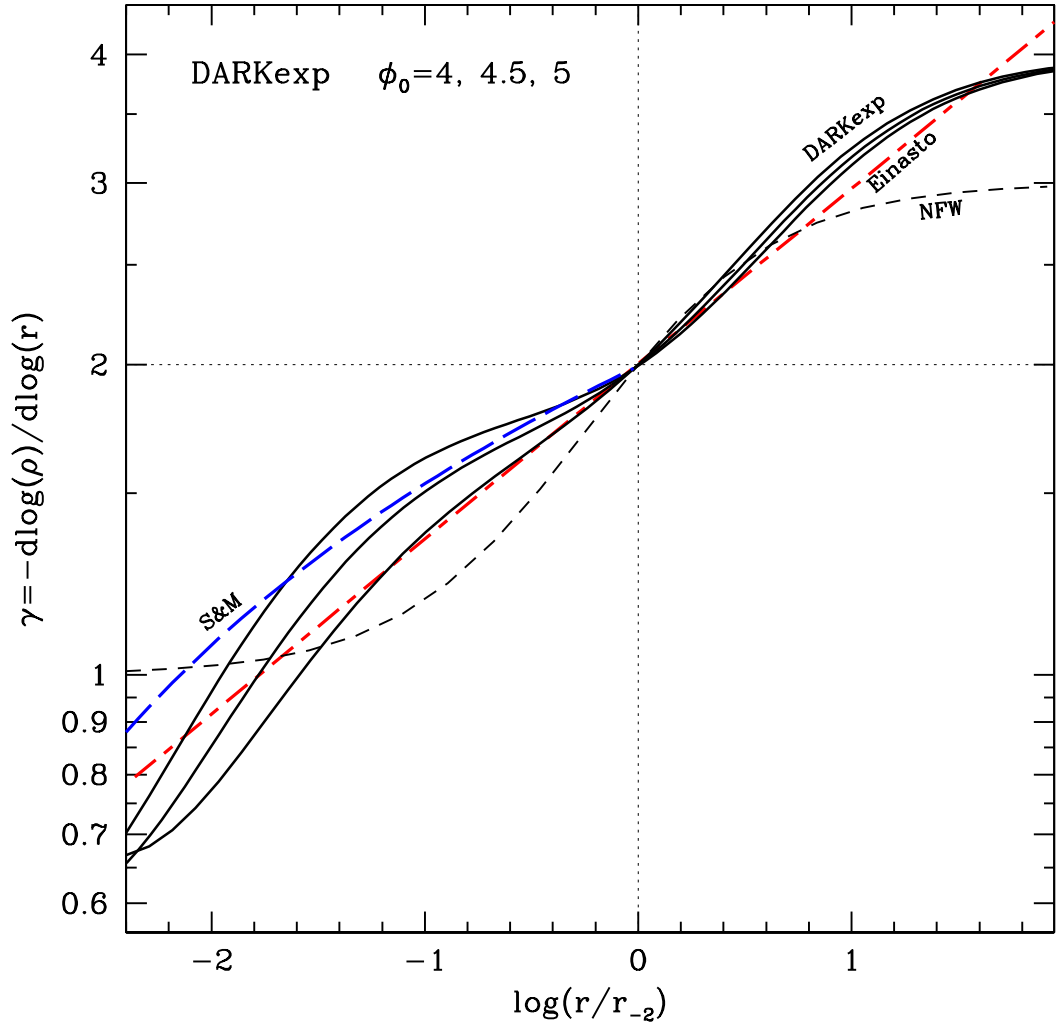}
\vskip-1.5in
\caption{Comparison of DARKexp density profile slopes (three solid black curves) 
with N-body fitting functions: Einasto (\cite{nav04} $\alpha=0.17$, short-long-dash red 
straight line), NFW (short-dash black curve), and S\&M (\cite{sm09}; long-dash blue 
curve). The last profile is meant to fit only the inner region of N-body halos, 
so the radial range outside of $r_{-2}$ is not plotted.} 
\label{slopeslim}
\end{figure}

As in Paper II, one can compare the density profiles of numerically generated halos
to those of isotropic DARKexp models. In Figure~\ref{slopeslim} we plot the
density slope $\gamma$ vs.\ the log of the radius. The original NFW fitting formula 
(short-dash black curve) is fairly similar to DARKexp (three solid back curves). 
The Einasto profile (short-long-dash red straight line) tracks DARKexp considerably
better than NFW, and does so for more than three decades in radius. 
While the Einasto is a straight line in the figure, DARKexp is `concave', i.e.,
its density slope changes less rapidly with increasing radius. The S\&M profile 
(long-dash blue curve) also has some concavity, and in this sense, it curves 
to mimic the DARKexp in the inner halo region. It is interesting to note that 
as the successive fitting formulae become better descriptors of the N-body halos
(NFW $\rightarrow$ Einasto $\rightarrow$ S\&M) they also get closer to the shape 
of the DARKexp density profile.

Since DARKexp ($\phi_0\approx 4-5$) and N-body density profiles are very similar, and 
we already showed in Section~\ref{ani2} that anisotropy can be ignored for our 
limited purposes, we can assume isotropy and obtain $N(E)$ from the density fitting 
functions of N-body simulations. From $\rho(r)$ we first calculate $f(E)$, using the 
isotropic Eddington formula, and then $N(E)$\footnote{
One might be tempted to use the isotropic Jeans equation, thereby obtaining the velocity 
dispersion profile $\sigma(r)$, and then estimating kinetic energy as $\frac{1}{2}\sigma^2$. 
Combining this with the potential gives $E$ for particles at that radius. This procedure 
will give wrong results because the velocity distribution function (VDF) is implicitly 
assumed to be Maxwellian, which is wrong. The importance of VDF and its deviations from 
Maxwellian was discussed in \cite{k04} and \cite{h09}.}.

\begin{figure}[t]
\epsscale{0.95}
\plotone{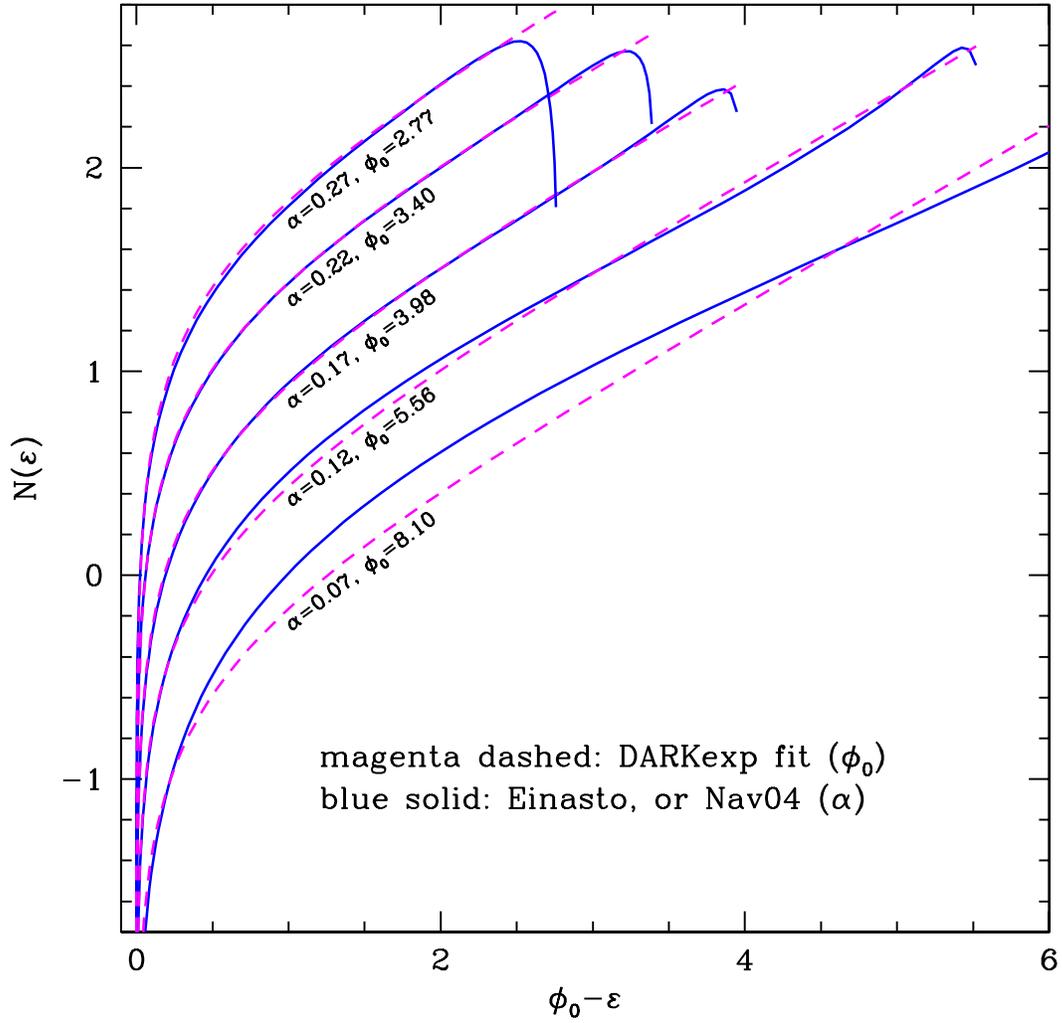}
\vskip-1.5in
\caption{Comparison between N(E) for DARKexp and N-body halos; see 
Section~\ref{compA}. A range of Einasto profiles (solid blue), with 
$\alpha=0.07, 0.12, 0.17, 0.22, 0.27$ were fit with DARKexp (dashed magenta), 
and the fitted $\phi_0$ values are shown in the plot. The best match occurs 
for the $\alpha$ value ($\approx 0.17$) which happens to be the average of 
simulated halos \citep{nav04}. For $\alpha\simlt 0.1$ the two 
distributions do not fit at all, and these $\alpha$ values are not encountered 
in simulations.} 
\label{rhotoNE_N04}
\end{figure}

The results are shown in Figure~\ref{rhotoNE_N04} for five Einasto profiles (solid 
blue curves), parametrized by the $\alpha$ index of \cite{nav04}; $\alpha$ between 
0.07 and 0.27, in steps of 0.05. In calculating these we used 7 radial decades of $\rho(r)$
between $10^{-5}\;r_{-2}$ and  $100\;r_{-2}$, which contain most of the systems' mass. 
We then fitted Einasto $N(E)$ with the DARKexp form, $N/N_0=\exp[-\beta_T(E-\Phi_0)]-1$.
The fitting yields $N_0$ and $\beta_T$.  As before, the product $\beta_T \Phi_0$ is 
the dimensionless potential depth $\phi_0$, the only shape parameter in DARKexp. 

In Figure~\ref{rhotoNE_N04} the DARKexp fits are represented by dashed magenta curves, 
and the best fit $\phi_0$ are indicated in the plot. Among these five cases the best 
match occurs at $\alpha\approx 0.17$, which is also the average of halos simulated 
by \cite{nav04}. For $\alpha\simlt 0.1$ the two distributions do not fit at all; and, 
perhaps not coincidentally, such $\alpha$ values are not encountered in simulations.

\section{Conclusions}

It is often said that $N(E)$ is primarily determined by $\rho(r)$, and visa 
versa \citep{BT87,mtj89}, and that $\beta(r)$ has very little 
effect on this. We have quantified this statement for the specific case of 
DARKexp $N(E)$ and $\rho(r)$. We find that if DARKexp halos had $\beta(r)$ 
shapes similar to those of N-body simulated halos, their density and energy 
distributions could not be distinguished from those of isotropic DARKexp 
halos. This means that even without a full theory of collisionless equilibrium
$N(E,L)$, which would presumably explain $\beta(r)$ of simulated halos, one can 
compare the DARKexp and N-body $N(E)$, while ignoring non-zero anisotropy. 
We have carried out this comparison in two ways, using (1) the actual energy 
distribution extracted from simulations, and (2) N-body $\rho(r)$ fitting 
functions, and $N(E)$ computed from $\rho(r)$ using the isotropic Eddington 
formula. Both of these methods agree that DARKexp $N(E)$ with $\phi_0\approx 4-5$ 
is an excellent match to N-body $N(E)$. This suggests that statistical mechanical 
principles of maximum entropy can be used to explain the equilibrated final 
product of N-body simulations.
In the future we will extend the maximum entropy principle used to derive 
DARKexp to include $L$. This will hopefully explain $\beta(r)$ observed in 
simulated halos. 

\acknowledgments
The Dark Cosmology Centre is funded by the Danish National Research Foundation.
LLRW would like to thank the hospitality of the Dark Cosmology Centre and the 
University of Z\"urich where part of this work was carried out.
The authors are grateful to Stefan Gottl\"ober, who kindly agreed for one of 
his CLUES simulations, ({\tt {http://www.clues-project.org/simulations.html}})
to be used in this paper. The simulation has been performed at the Leibniz
Rechenzentrum (LRZ) Munich.

{}


\begin{thebibliography}{}

\bibitem[An \& Evans(2006)]{ae06}
An, J. H. \& Evans, N. W. 2006. ApJ, 642, 752

\bibitem[Binney \& Tremaine(1987)]{BT87}
Binney, J. \& Tremaine, S. 1987, {\it Galactic Dynamics}, Princeton University Press,
1st edition

\bibitem[Ciotti \& Morganti(2010)]{cm10}
Ciotti, L. \& Lucia Morganti, L. 2010, Preprint, arXiv:1006.2344.

\bibitem[Efthymiopoulos et al.(2007)]{e07}     
Efthymiopoulos, C., Voglis, N. \& Kalapotharakos, C. 2007, 
in {\it Topics in Gravitational Dynamics}, Lecture Notes in Physics, Vol 729, 
Springer Berlin/Heidelberg. 
(Link: {http://www.springerlink.com/content/3401462523t3076h})

\bibitem[Hansen et al.(2010)]{h10}
Hansen, S. H., Juncher, D. \& Sparre, M. 2010, Preprint, arXiv:1005.1643 

\bibitem[Hansen(2009)]{h09}
Hansen, S. H. 2009, 694, 1250

\bibitem[Hansen \& Moore(2006)]{hm06}
Hansen, S. H. \& Moore, B. 2006, NewA, 11, 333

\bibitem[H\'enon(1959)]{h59}
H\'enon, M. 1959, Ann. d'Astrophys. 22, 126

\bibitem[H\'enon(1973)]{h73}
H\'enon, M. 1973, A\&A. 24, 229

\bibitem[Hjorth \& Williams(2010)]{hw10}
Hjorth, J. \& Williams, L. L. R. 2010, ApJ, in press (Paper I)

\bibitem[Kazantzidis et al.(2004)]{k04}
Kazantzidis, S., Magorrian, J. \& Moore, B. 2004, ApJ, 601, 37

\bibitem[Merritt et al.(1989)]{mtj89}
Merritt, D., Tremaine, S. \& Johnstone, D. 1989, MNRAS, 236, 829	

\bibitem[Merritt et al.(2006)]{m06}
Merritt, D., Graham, A. W., Moore, B., Diemand, J., \& Terzic, B. 2006, AJ, 132, 2685

\bibitem[Navarro et al.(1997)]{nfw97}
Navarro, J. F., Frenk, C. S. \& White, S. D. M. 1997, ApJ, 490, 493

\bibitem[Navarro et al.(2004)]{nav04}
Navarro, J. F. et al. 2004, MNRAS, 349, 1039

\bibitem[Navarro et al.(2010)]{nav10}
Navarro, J. F. et al. 2010, MNRAS, 402, 21

\bibitem[Schwarzschild(1979)]{sch79}
Schwarzschild, M. 1979, ApJ, 232, 236

\bibitem[Stadel et al.(2009)]{sm09}
Stadel, J., Potter, D., Moore, B., Diemand, J., Madau, P., Zemp, M., Kuhlen, M. \& Quilis, V.
2009, MNRAS, 398, L21 

\bibitem[Williams \& Hjorth(2010)]{wh10}
Williams, L. L. R. \& Hjorth, J. 2010, ApJ, in press (Paper II)

\bibitem[Wojtak et al.(2008)]{w08}
Wojtak, R., Lokas, E. L., Mamon, G. A., Gottloeber, S., Klypin, A. \& Hoffman, Y.
2008, MNRAS, 388, 815

\end{thebibliography}
\end{document}